\documentclass[prl,aps,amsmath,floatfix,nofootinbib,twocolumn]{revtex4-2}

\usepackage{latexsym}
\usepackage{graphicx}

\usepackage{hyperref}

\def\bibi{\bibitem}

\def\ttl#1{{\it #1}}

\def\d{\delta}
\def\e{\epsilon}                
\def\f{\phi}                    
\def\g{\gamma}

\def\j{\psi}

\def\m{\mu}
\def\n{\nu}

\def\p{\pi}                     

\def\J{\Psi}

\def\P{\Pi}

\def\cl{{\cal L}}

\def\co{{\cal O}}

\def\cbo{{\,\raise-.15ex\Sc [\,}}                       

\def\sl#1{\rlap{\hbox{$\mskip 1 mu /$}}#1}      
\def\Sl#1{\rlap{\hbox{$\mskip 3 mu /$}}#1}      

\def\svev#1{\left\langle #1\right\rangle}       

\def\ddt#1{{\buildrel {\hbox{\LARGE .\kern-2pt.}} \over {#1}}}

\def\eg{\mbox{\it e.g.}}

\def\tr{{\rm tr}\,}

\def\half{{1\over 2}}

\def\det{{\rm det}}

\def\mysec#1{\bigskip \noindent {\em #1.\ }}

\def\bj{\overline\psi}

\def\textit#1{{\it \!\!\! #1 \!\!}}

\def\invsl{\frac{1}{\sl{\partial}}}

\begin{document}

\title{Propagator zeros and lattice chiral gauge theories}

\author{Maarten Golterman}
\affiliation{Department of Physics and Astronomy,
  San Francisco State University, San Francisco, CA 94132, USA}

\author{Yigal Shamir}
\affiliation{Raymond and Beverly Sackler School of Physics and Astronomy,
Tel~Aviv University, 69978, Tel~Aviv, Israel}

\begin{abstract}
Symmetric mass generation (SMG) has been advocated as a mechanism
to render mirror fermions massive without symmetry breaking,
ultimately aiming for the construction
of lattice chiral gauge theories.  It has been argued that in an SMG phase,
the poles in the mirror fermion propagators are replaced by zeros.
Using an effective lagrangian approach,
we investigate the role of propagator zeros when the gauge field is turned on,
finding that they act as coupled ghost states.
In four dimensions, a propagator zero makes
an opposite-sign contribution to the one-loop beta function
as compared to a normal fermion.  In two dimensional abelian theories,
a propagator zero makes a negative contribution to the photon mass squared.
In addition, propagator zeros
generate the same anomaly as propagator poles.
Thus, gauge invariance will always be maintained in an SMG phase,
in fact, even if the target chiral gauge theory is anomalous,
but unitarity of the gauge theory is lost.
\end{abstract}

\maketitle

\mysec{Introduction}
Chiral gauge theories play an important role in particle physics.
The Standard Model is a chiral gauge theory,
and many theoretical models for physics beyond the Standard Model
are based on chiral gauge theories as well.   At present,
no nonperturbative gauge invariant definition
of (anomaly free) nonabelian chiral gauge theories exists.
Unlike the case of vector-like asymptotically free gauge theories
such as QCD, for which the lattice provides a nonperturbative definition,
the physics of chiral gauge theories is understood mostly
in a perturbative framework, supplemented by incomplete and rather qualitative
arguments based on large-$N$ expansions, anomaly matching,
duality relations, and other phenomenological approaches.

The nonperturbative construction of chiral gauge theories
on the lattice is a long-standing challenge,
because of the fermion species-doubling problem.
The deep reasons underlying species doubling
were first discussed by Karsten and Smit \cite{KS},
tying the phenomenon to the chiral anomaly,
and then generalized by Nielsen and Ninomiya \cite{NN}.
In its simplest form, the doubling problem arises when
one considers a local free lattice hamiltonian in one spatial dimension.
The dispersion relation near $p=0$ will be $E=+p$ for one chirality.
Because of the periodicity of the Brillouin zone, unless there is
a non-analyticity in the dispersion relation,
there must exist another point $p_c$ such that, for small $\d p = p-p_c $,
the dispersion relation is $E=- \d p$, which signifies
a fermion of the opposite chirality, the doubler.

During the 80s and 90s there was much activity in this field, which lead to
a better understanding of the fundamental obstacles,
as well as to some successes.  For reviews, see Refs.~\cite{YSrev,MGrev}.
Building on the Ginsparg--Wilson relation \cite{GW} as well as on earlier work
by Kaplan \cite{Kaplan} and by Narayanan and Neuberger
\cite{NNchiralPRL,NNchiralNPB,HNoverlap,HNGW},
L\"uscher achieved the construction
of (anomaly-free) abelian chiral gauge theories \cite{MLabelian}.
He also constructed nonabelian chiral gauge theories
to all orders in lattice perturbation theory \cite{MLnonabelian,MLpertth}.
An alternative approach, where the chiral gauge invariance
is explicitly broken on the lattice, and is only restored
in the continuum limit, is the gauge-fixing approach.
The inclusion of a suitable gauge-fixing lattice action ensures
the existence of a novel critical point, where the target chiral gauge theory
emerges in the continuum limit
\cite{YS95,GFaction,BGSPRL,BGSlat97b,FNV,nachgt}.
For another approach, see Refs.~\cite{GK,Kd}.
While these proposals are based on nontrivial insights into the nature of
the problem, it is still an open question whether any of them will lead to a
complete nonperturbative definition of asymptotically free
chiral gauge theories on the lattice.

The last decade saw renewed interest in the
{\em mirror fermion} approach.  One starts from a vector-like fermion spectrum
containing both LH (left-handed) and RH (right-handed) fields.
For each irreducible representation of the gauge group,
the fermions of one chirality are included in
the target chiral gauge theory,
while the fermion fields of the opposite chirality
are unwanted ``mirror'' fermions.  Originally,
it was proposed to take the continuum limit in the broken (Higgs) phase
\cite{Montvay,Montvayrev}.  However, this approach does not allow
for full decoupling on the mirrors. The reason is that full decoupling
requires the Higgs vacuum expectation value $v$ to diverge
in physical units when taking the continuum limit.
But since the mass of the gauge bosons in the Higgs phase is $\sim gv$,
this would imply that the gauge bosons decouple as well.
Keeping $v$ physical instead would imply that mirrors are also physical.
An interesting question is whether such an approach is phenomenologically
viable, but the underlying gauge theory remains vector-like, and not chiral.

Since then, the focus has shifted to attempts to decouple the mirror fermions
using some strong non-gauge interaction while at the same time
keeping the chiral gauge symmetry unbroken.  In principle,
this would allow for a complete decoupling, with the mirrors
obtaining a mass of the order of the ultraviolet (lattice) cutoff of the theory.
For a recent review, see Ref.~\cite{mirrorrev}.  In an asymptotically free theory,
the gauge interaction itself is controlled by the gaussian fixed point,
and turning it on or off is not expected to
change the elementary fermion spectrum.
The elementary fermion spectrum is thus controlled by
the {\em reduced model}, obtained by turning off the gauge interaction.
The reduced model contains the fermion fields and, possibly,
additional scalar fields.  The original (target) chiral gauge symmetry $G$
turns into a global symmetry of the reduced model.

The dynamical question is what the phase diagram of the reduced model
looks like \cite{YSrev,MGrev,mirrorrev}.
There is always a free-fermion limit containing both LH and RH
massless fermions in the same (in general, reducible) representation
of the symmetry group $G$ to be gauged,
which constitutes a vector-like spectrum.
By turning on specially designed multi-fermion or Yukawa interactions,
one hopes to achieve {\em symmetric mass generation} (SMG).
In this Letter, we use the term SMG for a strong-coupling phase in the
reduced model where
(a) the mirror fermions develop a mass gap of the order of the lattice cutoff,
(b) the target chiral fermions remain massless,
and (c) the symmetry $G$ is unbroken.
The low-energy limit then consists of a chiral fermion spectrum
in the desired representation of the (unbroken) symmetry $G$.
If the SMG paradigm is successful, one would hope to recover
the target chiral gauge theory when the gauge field is turned back on.

Starting from the Smit--Swift \cite{SS} and Eichten--Preskill \cite{EP} models,
many unsuccessful attempts were made over the years to find an SMG phase,
and the dynamical reasons underlying this failure were investigated
\cite{GPS,BDS,GPR}.
The last decade or so has seen renewed interest in the {\em 3450 model}.
This is an anomaly free, two-dimensional abelian chiral gauge theory
containing fermions with charges 3 and 4 of one handedness,
and fermions with charges 5 and 0 of the opposite handedness.\footnote{
The zero charge fermion does not couple to the gauge field,
but it participates in the multi-fermion interactions that gap the mirrors
in the reduced model.
}
The focus on
two-dimensional models is motivated by the relative simplicity of gauge theories
in two dimensions, with the hope that lessons learned generalize to
the more interesting case of four-dimensional chiral gauge theories.

An attempt to gap the mirrors in the reduced version of the 3450 model
was made in Refs.~\cite{GPopp,CGP}.  The vacuum polarization was calculated
numerically, but the result indicated that the mirrors did not decouple.
A possible reason for this failure was discussed in Ref.~\cite{Yoshio}.
Recently, building on developments in condensed matter physics
\cite{SMGreview}, an SMG phase was reported in a different reduced version
of the 3450 model, with the SMG phase induced by especially designed
multi-fermion interactions \cite{WWPRB,DMW,3450PRL}.

The availability of a concrete construction allows us to reexamine the question
of whether the desired chiral gauge theory will indeed be recovered
when the gauge interaction is turned back on.  The question is nontrivial
for the following reason.  As already noted, the desired low-energy limit
of the reduced model is a theory of free, undoubled, massless chiral fermions.
This is essentially the domain of the no-go theorems.  Indeed,
it was argued in Ref.~\cite{YS93} that, under some very general assumptions,
the Nielsen--Ninomiya theorem will be applicable to an effective
low-energy hamiltonian of an unspecified underlying theory,
thus excluding a chiral spectrum in any reduced model
unless some of these assumptions are not satisfied.\footnote{%
  As explained in Ref.~\cite{BGSlat97b}, the gauge-fixing approach of
  Refs.~\cite{YS95,GFaction,nachgt} evades the generalized
  no-go theorem \cite{YS93}. See also Ref.~\cite{BGSPRL}.
}

A caveat, already noted in Ref.~\cite{YS93}, is that the fermion propagator may
contain a zero in the mirror channel. Equivalently, the effective hamiltonian
associated with the inverse propagator has a mirror pole,
and thus is nonlocal. While to our knowledge
the issue was not directly investigated in the SMG phase of the 3450 model,
there are strong general arguments that an SMG phase will always
be accompanied by the appearance of a zero in the propagator
\cite{SMGreview,YWOX,SYX,XX,CB,YHXV,LZY}.
This zero takes the place of the original massless mirror pole,
and its essential features are captured by the phenomenological expression
\begin{equation}
\label{pzero}
P_{R,L}\,\frac{i\sl{p}}{p^2+m^2} \ ,
\end{equation}
for the mirror fermion propagator valid near $p=0$,
where $P_{R,L}=\half(1\pm\g_5)$ are the chiral projectors.
The large mass scale $m$ characterizes the mass gap in the mirror sector
after SMG has taken place. In the generic case, it is
presumably on the order of the lattice cutoff.

The goal of this Letter is to examine the consequences
of a propagator zero of a massive charged fermion in the reduced model
when the gauge field is turned back on. We will first discuss
the contribution of a propagator zero to the vacuum polarization,
and then its impact on anomalies, in both two and four dimensions.

\mysec{Propagator zero and the vacuum polarization}
Before discussing the physics of a propagator zero,
let us briefly recall the calculation of the vacuum polarization
for normal RH or LH fields.  Starting from the gauge invariant lagrangian
(in euclidean space)
\begin{equation}
\label{LRH}
\cl = \bj ({\sl{\partial}}+ig\Sl{A}) P_{R,L} \j \ ,
\end{equation}
the vacuum polarization diagram is
\begin{eqnarray}
\P^{R,L}
&=& \half \svev{\left(-ig\, \bj \Sl{A} P_{R,L} \j\right)^2}
= \frac{g^2}{2}\, \tr\!\left(
\invsl \Sl{A} \invsl \Sl{A} P_{R,L}\right)
\nonumber\\
&=& \P^e \pm \P^o \ ,
\label{RLvacpol}
\end{eqnarray}
where $\P^e$ and $\P^o$ are the parity-even and parity-odd parts.
In $d=4$ the parity-odd part vanishes, whereas in $d=2$ it gives rise
to the anomaly, which we discuss later.

Moving on,
we consider a RH fermion field $\j_R = P_R \J_R$, $\bj_R = \bj_R P_L$,
for which the propagator pole has been replaced by a propagator zero.
The momentum space RH propagator takes the form (compare Eq.~(\ref{pzero}))
\begin{equation}
\label{propR}
\int d^dx\,e^{-ipx} \svev{\j_R(x)\bj_R(0)}
= P_R\, \frac{i\sl{p}}{m^2} \Big(1+\co(p^2/m^2)\Big)\ .
\end{equation}
The (even) dimension of spacetime is $d$.
For simplicity, we assumed that the propagator zero occurs at $p=0$.
Our results are unchanged if the zero occurs
at a different location in the Brillouin zone.
The corrections to the leading behavior are suppressed by
powers of $p^2/m^2$ near the zero, and do not affect our results either.%
\footnote{We assume that the (euclidean) lattice theory regains
full rotational invariance at large distances,
and thus that the zeros of the propagator are relativistic.
Relaxing this assumption is likely to lead to yet
worse problems than those we find in this Letter.}

Keeping only the leading behavior, in operator language
the RH propagator is
\begin{equation}
\label{propRx}
\svev{\j_R\,\bj_R} = P_R\, \frac{\sl\partial}{m^2}\ ,
\end{equation}
and the corresponding free effective lagrangian has a RH pole,
\begin{equation}
\label{L0}
\cl_0 = m^2\, \bj_R \,\invsl\, \j_R \ .
\end{equation}
This effective lagrangian is nonlocal.
As we will see, this has consequences for the gauged theory at low energy,
even though the mirror fermion has a mass gap.

We introduce the gauge field via minimal coupling, as usual.
The effective lagrangian becomes
\begin{eqnarray}
\cl &=& m^2\, \bj_R\, \frac{1}{{\sl{\partial}}+ig\Sl{A}} \,\j_R
\label{EFTR}
\\
&=& m^2\, \bj_R \left( \invsl -ig \invsl \Sl{A} \invsl
-g^2 \invsl \Sl{A} \invsl \Sl{A} \invsl + \cdots \right) \j_R
\nonumber
\\
&\equiv& \cl_0 + \cl_{\rm int} \ .
\nonumber
\end{eqnarray}
The second line effectively defines the inverse
of ${{\sl{\partial}}+ig\Sl{A}}$ in terms of its expansion
in the coupling constant $g$, or equivalently,
in the external gauge field.\footnote{
We treat the gauge field perturbatively, and thus the operator
${\sl{\partial}}+ig\Sl{A}$ cannot have zero modes.}
The vacuum polarization now consists of a bubble and a tadpole diagram.
We start with the bubble diagram.
Expanding $\exp(-\int d^dx \cl_{\rm int})$ to second order
and using Eq.~(\ref{propRx}) gives
\begin{equation}
\label{twovertices}
\half \svev{\left(igm^2\, \bj_R\, \invsl \Sl{A} \invsl \,\j_R\right)^2}
= \frac{g^2}{2}\, \tr\!\left(P_R \Sl{A} \invsl \Sl{A} \invsl \right)
= \P^L \ .
\end{equation}
The tadpole contribution arises from expanding
$\exp(-\int d^dx \cl_{\rm int})$ to first order,
\begin{equation}
\label{tadpole}
g^2 m^2 \svev{\bj_R\, \invsl \Sl{A} \invsl \Sl{A} \invsl \,\j_R}
= -g^2\,\tr\!\left(P_R \Sl{A} \frac{1}{\sl{\partial}}
\Sl{A} \frac{1}{\sl{\partial}} \right)
= -2\P^L \ .
\end{equation}
The vacuum polarization of a RH propagator zero is thus
\begin{equation}
\label{tot}
\P^L - 2\P^L = -\P^L = -\P^e + \P^o \ .
\end{equation}
Comparing with Eq.~(\ref{RLvacpol}) we see that the parity-even part
flips its sign (the parity-odd part remains the same).
A similar result is obtained for a LH propagator zero.

Equation~(\ref{tot}) is our main result.  To start,
it demonstrates that a propagator zero acts as a coupled state at low energy
in the gauge theory through the poles it generates in the vertices.
Let us concentrate on the parity-even part $\P^e$, postponing
the parity-odd part to our discussion of anomalies below.
Differentiating it twice with respect to $A_\m$ gives the familiar result
\begin{equation}
\label{vacpolmunu}
\P^e_{\m\n}(k) = -\frac{g^2}{2} \int\frac{d^dp}{(2\p)^d}\,
\frac{\tr({\sl{p}}\g_\m({\sl{p}}+{\sl{k}}) \g_\n )}{p^2(p+k)^2} \ .
\end{equation}
In the abelian case, this result is valid for a unit-charge field.
In the nonabelian case, the result is multiplied by $\tr(T_aT_b) = T \d_{ab}$,
where $T$ is the group trace in the fermion representation.

The integral~(\ref{vacpolmunu}) is UV divergent
and may be computed for example using dimensional regularization.
In four dimensions, after renormalization, the result is
\begin{equation}
\label{vacpollog}
\P^e_{\m\n} = \frac{g^2}{24\p^2} (k^2\d_{\m\n}-k_\m k_\n)
\log\left(\frac{k^2}{4\p\m^2}\right) + \cdots \ ,
\end{equation}
where $\m$ is the renormalization scale.  The dots stand for a finite term,
which is subleading compared to the logarithm.
The logarithm in Eq.~(\ref{vacpollog}) determines the fermion contribution
to the one-loop beta function.  Because $\P^e$ flips sign
between Eqs.~(\ref{RLvacpol}) and~(\ref{tot}), it follows that
the contribution of a propagator zero to the one-loop beta function
has the same absolute value as a normal fermion, but an opposite sign.
Similarly, the contribution of a propagator zero to the imaginary part
of the vacuum polarization will be negative,
in violation of the optical theorem.
The propagator zero thus acts as a ghost state, which ruins
the unitarity of the gauge theory in the SMG phase.

A similar result was found in the 80s in Refs.~\cite{Peli1,Peli2}, where
a specific nonlocal lattice action \cite{Rebbi} was studied in which
the fermion doublers were replaced by poles in the action,
or equivalently, zeros in the fermion propagator.
We have employed an effective lagrangian approach to generalize
this result, finding that the effect occurs in complete generality.

Let us illustrate the consequences of these findings for the
mirror fermion approach.  For definiteness, we start with
a four-dimensional lattice gauge theory of massless domain-wall fermions,
in which the LH spectrum on one wall constitutes the fermion spectrum
of the target chiral gauge theory.  The RH spectrum on the other wall
is the mirror spectrum.  Before we turn on any interactions to gap the mirrors,
the theory is vector-like.  Let the contribution of these Dirac fermions
to the coefficient of the one-loop beta function be $b_f$.
This contribution is split evenly
between the LH and RH fields, with each contributing $b_f/2$.
The one-loop contribution of the target chiral fermions is thus $b_f/2$.

Now let us assume that the RH fermions have been gapped in an SMG phase,
with each RH propagator pole replaced by a propagator zero.
What we find is that the total fermion contribution to the
one-loop beta function will now vanish, because the contribution
of the LH target chiral fermions, which is $b_f/2$, is cancelled
by that of the RH propagator zeros, which is $-b_f/2$.

In two dimensions, restricting ourselves to the abelian case for simplicity,
the vacuum polarization gives rise to a photon mass squared
$m_{\rm ph}^2$ which, in turn, is proportional to
the sum of the U(1) charges squared \cite{JS}
(for a textbook discussion, see Ref.~\cite{PS}).
The target chiral gauge theory, which consists of half of the massless
fermion degrees of freedom of the initial domain-wall fermion lattice theory
(and contains chiral fields of both chiralities),
is expected to generate a mass squared $m_{\rm ph}^2/2$.
Again, we find that the contribution of the propagator zeros
has the same absolute value and an opposite sign,
and is thus given by $-m_{\rm ph}^2/2$.
The result is a vanishing photon mass in the SMG phase.
These effects, in both four and two dimensions,
reflect the role of the propagator zeros as coupled ghost states.

\mysec{Propagator zero and anomalies}
We first briefly recall the calculation of the axial anomaly,
starting with the two dimensional case. One way to perform the calculation
is to introduce both a vector gauge field
$V_\m$ and an axial gauge field $A_\m$, and consider the lagrangian
\begin{equation}
\label{lagVA}
\cl = \bj\Big({\sl\partial} +\frac{ig}{2}({\sl{V}}+\Sl{A}\g_5)\Big)\j \ .
\end{equation}
If we set $V=A$, this reduces to
\begin{equation}
\label{lagR}
\cl = \bj ({\sl\partial} + ig\Sl{A} P_R ) \j \ .
\end{equation}
This lagrangian describes a RH fermion coupled to the gauge field
together with a free LH fermion.
In dimensional regularization Eq.~(\ref{lagVA}) is thus
an adequate starting point for the calculation of the anomaly of a RH fermion.
When calculating the anomaly using dimensional regularization,
one must make sure to use a consistent definition of $\g_5$,
or of its two-dimensional counterpart (see, for example, Refs.~\cite{PS,collins}).
Below, we will be careful to only perform manipulations which are
valid within dimensional regularization.  This means, for example,
that when we calculate a diagram we are allowed
to anticommute $\g_5$ through $\Sl{A}$,
because the external gauge field (and momenta)
are two- or four-dimensional by assumption.  However, we are not allowed
to anticommute $\g_5$ through the fermion propagator,
because of its evanescent part.

The two-dimensional axial anomaly arises from the bubble diagram with
one $V$ vertex and one $A$ vertex.  Starting from the lagrangian~(\ref{lagVA}),
this diagram is given by
\begin{equation}
\label{AVbubble}
\P^{AV} = \frac{g^2}{4}\, \tr\!\left(\invsl {\sl{V}} \invsl \Sl{A}\g_5\right)\ .
\end{equation}
In order to calculate the divergence of the axial current,
one replaces $\Sl{A}$ by ${\sl{q}}$, where $q$ is the external momentum.
The well-known result is $\propto \e_{\m\n} \partial_\m V_\n$.
Alternatively, using the two-dimensional relation
$\g_\m\g_5 \propto \e_{\m\n}\g_\n$ one can express the parity-odd part
of the vacuum polarization in terms of the parity-even part,
and then obtain the anomaly.  For details see, \eg, Ref.~\cite{PS}.

Proceeding to the case of a propagator zero, our starting point is
\begin{equation}
\label{zlagVA}
\cl = m^2\,
\bj\,\frac{1}{{\sl\partial} +\frac{ig}{2}({\sl{V}} + \g_5 \Sl{A})}\,\j \ .
\end{equation}
Note that this time $\g_5$ is to the left of $\Sl{A}$.
If we set $V=A$, we now obtain
\begin{eqnarray}
\label{zlagR}
\cl &=& m^2\, \bj\,\frac{1}{{\sl\partial} + ig P_R \Sl{A}}\,\j
\\
&=& m^2\left( \bj_R\,\frac{1}{{\sl\partial} + ig \Sl{A}}\,\j_R
+ \bj_L\,\frac{1}{\sl\partial}\,\j_L \right) \ .
\nonumber
\end{eqnarray}
The second equality holds in two and four dimensions,
where $\g_5$ anticommutes with the Dirac operator,
showing the adequacy of this starting point (compare Eq.~(\ref{EFTR})).
Again, for our perturbative calculations we define the inverses
of all differential operators that depend on external gauge fields
through their expansions in powers of $g$.

Like Eq.~(\ref{lagVA}), also Eq.~(\ref{zlagVA}) allows for the use of
dimensional regularization.
As in the case of the vacuum polarization,
the two-dimensional anomaly of a propagator zero involves
tadpole and bubble diagrams.
The details are similar, and the final result
is that the contribution of a propagator zero is $\P^{AV}$,
with a contribution $-\P^{AV}$ from the bubble diagram,\footnote{
  The minus sign comes from the location of $\g_5$ to the left of $\Sl{A}$
  in Eq.~(\ref{zlagVA}), instead of to the right in Eq.~(\ref{lagVA}).
}
and a contribution $+2\P^{AV}$ from the tadpole diagram.  In other words,
a propagator pole and a propagator zero give rise to the same anomaly.
This is consistent with our previous result,
that the parity-odd part of the vacuum polarization is the same
for a propagator pole and for a propagator zero.

The four-dimensional case is similar.
Starting from the lagrangian~(\ref{lagVA}),
the usual triangle diagram is given by
\begin{equation}
\label{AVV}
\P^{AVV} = -\frac{ig^3}{16}\, \tr\!\left(
\invsl {\sl{V}} \invsl {\sl{V}} \invsl \Sl{A}\g_5 \right) \ .
\end{equation}
For the contribution of a propagator zero, starting from
the nonlocal lagrangian~(\ref{zlagVA}) we have tadpole, bubble, and triangle
diagrams.  The triangle diagram for a propagator zero turns out to be
the same as for a propagator pole,\footnote{
  In this case the minus sign from the different placement of $\g_5$
  is canceled by the overall relative sign of the three vertices contributing
  to the triangle diagram.
}
while the bubble and tadpole diagrams cancel each other.
The final result is again that
a propagator pole and a propagator zero give rise to the same anomaly.

A formal but quick way to keep track of the relative signs is to note
that the fermion partition function defined by the lagrangian~(\ref{lagVA})
is $\det({\sl\partial} +\frac{ig}{2}({\sl{V}}+\Sl{A}\g_5))$, whereas the
partition function  defined by the nonlocal lagrangian~(\ref{zlagVA})
is $\det^{-1}({\sl\partial} +\frac{ig}{2}({\sl{V}}+\g_5\Sl{A}))$.
For the purpose of perturbative calculations,
the latter partition function may alternatively be expressed as a path integral
over a commuting complex field $\f$ carrying a spinor index, with lagrangian
$\f^\dagger ({\sl\partial} +\frac{ig}{2}({\sl{V}}+\g_5\Sl{A})) \f$.
It is straightforward to see that the diagrammatic expansion of
these determinants reproduces the relative signs we found above.

Let us revisit the mirror fermion approach to the construction of lattice
chiral gauge theories in the light of our findings.
As explained in the introduction, this approach was based on two premises.
The first premise is that gapping the mirrors in an SMG phase
will decouple them entirely from the low-energy gauge theory.
The second is that the target chiral fermion spectrum must be anomaly free,
in order to respect
the gauge invariance of the lattice theory after the mirrors have been gapped.

What we have found is that, if the mirror propagator poles are traded with
propagator zeros in the SMG phase, then they will remain coupled
to the gauge theory at low energy.  Gauge invariance is in fact
always maintained, regardless of whether or not the fermion spectrum
of the target chiral gauge theory is anomaly free.
The reason is that each propagator zero generates the same anomaly
as the corresponding propagator pole.  Thus, the propagator zeros
will always cancel the anomaly of the target chiral fermions,
as did the original mirror propagator poles.  The underlying reason
for this result is gauge invariance, which was exact before
the SMG dynamics was turned on, and remains so after.
In two dimensions, an interesting relation was found between SMG
and anomaly cancellation expressed through the ``boundary fully gapping rules''
of Ref.~\cite{WWPRB}, but it turns out that this plays no role in arriving
at our conclusions about anomaly cancellation between the target and
mirror sectors of a model with symmetric mass generation.

At the same time, unitarity is always lost, because the propagator zeros
contribute to the one-loop beta function in four dimensions,
or to the photon mass squared in two-dimensional abelian theories,
as ghost states.

In principle, apart from the propagator zeros
(and the target chiral fermions)
there could exist bound states of the lattice fields
that also contribute at large distances.
Such states, if present, would remedy the theory only if they
undo the effect of the propagator zeros
both for the beta function and for the anomaly,
without generating any other long distance effects.
It is hard to see how this would come about.

\vspace{1.5ex}
\noindent
{\em Acknowledgments.\ }
This work was initiated at the Aspen Center for Physics,
which is supported by National Science Foundation grant PHY-2210452.
YS wishes to thank the organizers and the participants of the workshop
``Emergent Phenomena of Strongly-Interacting Conformal Field Theories
and Beyond'' for stimulating discussions.  In particular, YS thanks Cenke Xu
and Yizhuang You for discussions and correspondence on symmetric mass generation.
YS also thanks the Department of Physics and Astronomy
at San Francisco State University for hospitality.
This material is based upon work supported by the U.S. Department of
Energy, Office of Science, Office of High Energy Physics,
Office of Basic Energy Sciences Energy Frontier Research Centers program
under Award Number DE-SC0013682 (MG).
YS is supported by the Israel Science Foundation under grant no.~1429/21.


\end{document}